\documentclass{article}
\usepackage{spconf,amsmath,graphicx,xcolor,url,booktabs,multirow,cite,subfig}

\newcommand{\parasection}[1]{\noindent \textbf{#1:}}

\title{jaCappella Corpus: A Japanese a Cappella Vocal Ensemble Corpus}
\name{Tomohiko Nakamura$^\dagger$, Shinnosuke Takamichi$^\dagger$, Naoko Tanji$^\dagger$, Satoru Fukayama$^\ddagger$, Hiroshi Saruwatari$^\dagger$\thanks{This work was supported by JSPS KAKENHI under Grants JP20K19818, JP21H04900, and JP21K12202, and NII CRIS collaborative research program operated by NII CRIS and LINE Corporation.}}
\address{
    $^\dagger$Graduate School of Information Science and Technology, The University of Tokyo\\
    7-3-1 Hongo, Bunnkyo-ku, Tokyo 113-8656, Japan \\
    $^\ddagger$National Institute of Advanced Industrial Science and Technology (AIST)\\
    2-4-7 Aomi, Koto, Tokyo 135-0064, Japan
}
\begin{document}
\ninept
\maketitle
\begin{abstract}
We construct a corpus of Japanese \emph{a cappella} vocal ensembles (\emph{jaCappella corpus}) for vocal ensemble separation and synthesis.
It consists of 35 copyright-cleared vocal ensemble songs and their audio recordings of individual voice parts.
These songs were arranged from out-of-copyright Japanese children's songs and have six voice parts (lead vocal, soprano, alto, tenor, bass, and vocal percussion).
They are divided into seven subsets, each of which features typical characteristics of a music genre such as jazz and enka.
The variety in genre and voice part match vocal ensembles recently widespread in social media services such as YouTube, although the main targets of conventional vocal ensemble datasets are choral singing made up of soprano, alto, tenor, and bass.
Experimental evaluation demonstrates that our corpus is a challenging resource for vocal ensemble separation.
Our corpus is available on our project page.
\end{abstract}
\section{Introduction}
\label{sec:intro}
Vocal ensemble is a widespread group singing form across cultures and languages and has been gathering attention in the field of music information retrieval (MIR) \cite{Cuesta2018ICMPC,Rosenzweig2020TISMIR_1,Cuesta2022PhD,Cuesta2020ISMIR,Sarkar2021Interspeech,Gover2020ISMIR,Petermann2020ISMIR,Mcleod2017AS,Chandna2022FSP,Tamaru2020IEICETIS}.
Owing to successes of deep learning in the this field \cite{Purwins2019IEEEJSTSP,Briot2020book}, recent studies on vocal ensemble have adopted a data-driven approach: for example, multipitch estimation \cite{Cuesta2022PhD,Cuesta2020ISMIR}, vocal ensemble separation \cite{Gover2020ISMIR,Petermann2020ISMIR,Sarkar2021Interspeech}, automatic transcription \cite{Mcleod2017AS}, synthesis of unison voices \cite{Chandna2022FSP}, and double-tracking \cite{Tamaru2020IEICETIS}.
For the development of such methods, the availability of suitable vocal ensemble datasets is essential.

There are a few publicly available datasets on vocal ensembles \cite{Cuesta2018ICMPC,Sarkar2021Interspeech,Cuesta2022PhD,Rosenzweig2020TISMIR_1}.
Table~\ref{tab:comp} shows the specifications of the vocal ensemble datasets.
Most of them focus on choral singing, particularly four-part singing of soprano (S), alto (A), tenor (T), and bass (Bs).
This SATB format is widespread in Western music and audio recordings of traditional choral music have been dealt with.
Some of the datasets \cite{Cuesta2022PhD,Rosenzweig2020TISMIR_1} include recordings of vocal ensembles with multiple singers per voice part to analyze interactions between singers, while most vocal ensemble studies focus on vocal ensembles with one singer per voice part \cite{Mcleod2017AS,Cuesta2020ISMIR,Cuesta2022PhD,Gover2020ISMIR,Petermann2020ISMIR,Sarkar2021Interspeech}.
The datasets of \cite{Cuesta2022PhD,Rosenzweig2020TISMIR_1} include audio recordings of singing voices during exercises and a few songs were often used.
In \cite{Sarkar2021Interspeech}, more than 40 songs were used to develop a vocal ensemble separation method.
This should be because the use of same songs for training and test leads to data leakage.

Unlike choral singing, recent vocal ensemble songs include more various voice parts such as lead vocal (Vo) and vocal percussion (VP).
VP is the imitation of percussion sounds by mouth and can accounts for rhythm.
Thus, it fits the vocal ensemble arrangement of various music genre songs such as jazz and reggae.
Such vocal ensemble songs have been actively shared in social media services (e.g., YouTube and TikTok), forming a new vocal ensemble style.
Despite their importance, there do not exist datasets that feature this style and can be easily used for research, to the best of our knowledge.
Since the availability of data is crucial for developing MIR methods on vocal ensembles, the lack of suitable datasets requires a costly and painstaking process of data collection.

In this paper, we construct a corpus of Japanese a cappella vocal ensembles, named \emph{jaCappella corpus}, by arranging out-of-copyright Japanese children's songs.
It includes 35 vocal ensemble songs composed of seven subsets (five songs per subset).
The songs in each subset were arranged with a different music genre, which ensures the variety in music genre and singing styles.
All songs have six voice parts: Vo, S, A, T, Bs, and VP.
This format is popular in Japanese vocal ensemble arrangement, which we examined by collecting commercial vocal ensemble songs.
To avoid copyright-related restriction, we obtained all necessary copyrights and neighboring rights of our songs.
Reserving these rights allows users of the jaCappella corpus to share processed audio signals to the extent necessary for research and it can also open the way for commercial use.
This is one of the advantages of manually creating music data because ensuring such a broad use is usually difficult to achieve with a method that artificially mixes singing voices from existing datasets~\cite{Cuesta2020ISMIR}. 
To further enhance the usability of this corpus, we provide the musical scores of the songs in the MusicXML format~\cite{musicxmlbook}.

Our corpus includes audio signals of individual voice parts sung by 20 semi-professional singers.
A vocal ensemble was assembled per subset and each voice part was sung by one singer.
The recordings were performed per singer.
We can use the recorded signals for various tasks on vocal ensembles.
For example, we can use the jaCappella corpus for vocal ensemble separation, which we will demonstrate in Section~\ref{sec:exp}.
In this paper, we focus on monaural mixtures but our corpus can be used for multichannel situations such as bleeding sound reduction (e.g., \cite{Mizobuchi2021APSIPA}).
Another task of our interest is singing voice synthesis for vocal ensemble, which we refer to as vocal ensemble synthesis.
This research direction aims to clarify how to artificially reproduce vocal ensemble performed by humans.
Since our corpus covers various music genres, it would be useful to synthesize ensemble singing voices with singing styles other than choral singing.
The corpus is available at \url{https://tomohikonakamura.github.io/jaCappella_corpus/}.

\begin{table*}[t]
    \centering
    \caption{Specifications of our jaCappella corpus and conventional vocal ensemble datasets}
    {
    \centering
    \setlength{\tabcolsep}{5pt}
    \begin{tabular}{c|cccccc} \toprule
        Corpus/Dataset & Voice parts & Duration [min] & \# songs & Genre & Publicly avail. \\
        \midrule
        Choral Singing \cite{Cuesta2018ICMPC} & S, A, T, Bs & 7 & 3 &Choral music & Yes \\
        Dagstuhl ChiorSet \cite{Rosenzweig2020TISMIR_1} & S, A, T, Bs & 55 & 2 & Choral music & Yes \\
        ESMUC Choir \cite{Cuesta2022PhD} & S, A, T, Bs & 31 & 3 & Choral music & Yes \\
        Bach Chorales and & \multirow{2}{*}{S, A, T, Bs} & \multirow{2}{*}{104} & \multirow{2}{*}{48} & \multirow{2}{*}{Choral and barbershop music} & \multirow{2}{*}{No} \\
        Barbershop Quartets \cite{Sarkar2021Interspeech} & & & & & \\
        \multirow{2}{*}{\textbf{jaCappella (ours)}} & \multirow{2}{*}{Vo, S, A, T, Bs, VP} & \multirow{2}{*}{34} & \multirow{2}{*}{35} & Jazz, punk rock, bossa nova, popular,& \multirow{2}{*}{Yes} \\
        & & & & reggae, enka, neutral (children's song) & \\
        \bottomrule
    \end{tabular}
    }
    \label{tab:comp}
\end{table*}

\section{Our jaCappella Corpus}
We constructed the jaCappella corpus in three steps: corpus design, arrangement into vocal ensemble songs, and recording of singing voices.
In the following, we describe the details of the three steps and analyze the distributions of lyrics, comparing the created songs with commercial vocal ensemble songs.

\subsection{Corpus design} \label{sec:design}
\begin{table}[t]
    \centering
    {
        \caption{Gender frequency of commercial music collection}
        \label{tab:gender_frequency}
        \begin{tabular}{c|cccccc}
            \toprule
            \multirow{2}{*}{Gender} & \multicolumn{6}{c}{Voice part [\%]} \\
            & Vo & S & A & T & Bs & VP \\
            \midrule
            Male & 81.0 & 0.0 & 0.0 & 66.7 & 100.0 & 100.0 \\
            Female & 19.0 & 100.0 & 100.0 & 33.3 & 0.0 & 0.0 \\ \bottomrule
        \end{tabular}
    }
\end{table}
We design the jaCappella corpus based on analysis of commercial vocal ensemble songs.
We collected 60 musical pieces from a Japanese sheet music store \cite{elevart} and examined them in terms of voice parts, gender of singers, and variety of music genres.
In addition, we address copyright issues to legally distribute our corpus.

\parasection{Voice parts}
46 of the 60 musical pieces have six voice parts.
42 of the 46 pieces consist of Vo, S, A, T, Bs, and VP, while the other 4 pieces have fourth chorus parts instead of Vo.
Following the majority, we adopted the six-part format of Vo, S, A, T, Bs, and VP.
We call the 42 pieces \emph{the commercial music collection} and determined other conditions on the basis of this collection.

\parasection{Gender of singers}
Table~\ref{tab:gender_frequency} shows the gender frequencies of the singers of the commercial music collection.
The female singers are dominant for S and A. The male singers are dominant for T, Bs, and VP.
We followed the majority for these parts.
Since the gender of a Vo singer strongly depends on musical pieces, we chose female Vo singers for convenience on singing voice synthesis.

\parasection{Variety of music genres}
The commercial music collection includes various music genres, for example, Japanese popular music, rock, and enka (Japanese traditional ballad).
The variety in genre is essentially different from choral songs and should be considered in corpus design.
We divided songs into several subsets and arranged the songs in each subset to different genres.

\parasection{Copyrights of musical pieces}
The distribution of a corpus that includes commercial musical pieces is extremely restrictive because we must not violate their copyrights and neighboring rights.
To avoid this restriction, we decided to arrange out-of-copyright songs into vocal ensemble songs and reserve their copyrights and neighboring rights necessary for research purposes.

\subsection{Arrangement into vocal ensemble songs}\label{sec:acapella_songs}
\begin{figure}[t]
    \centering
    \includegraphics[width=0.99\columnwidth,clip]{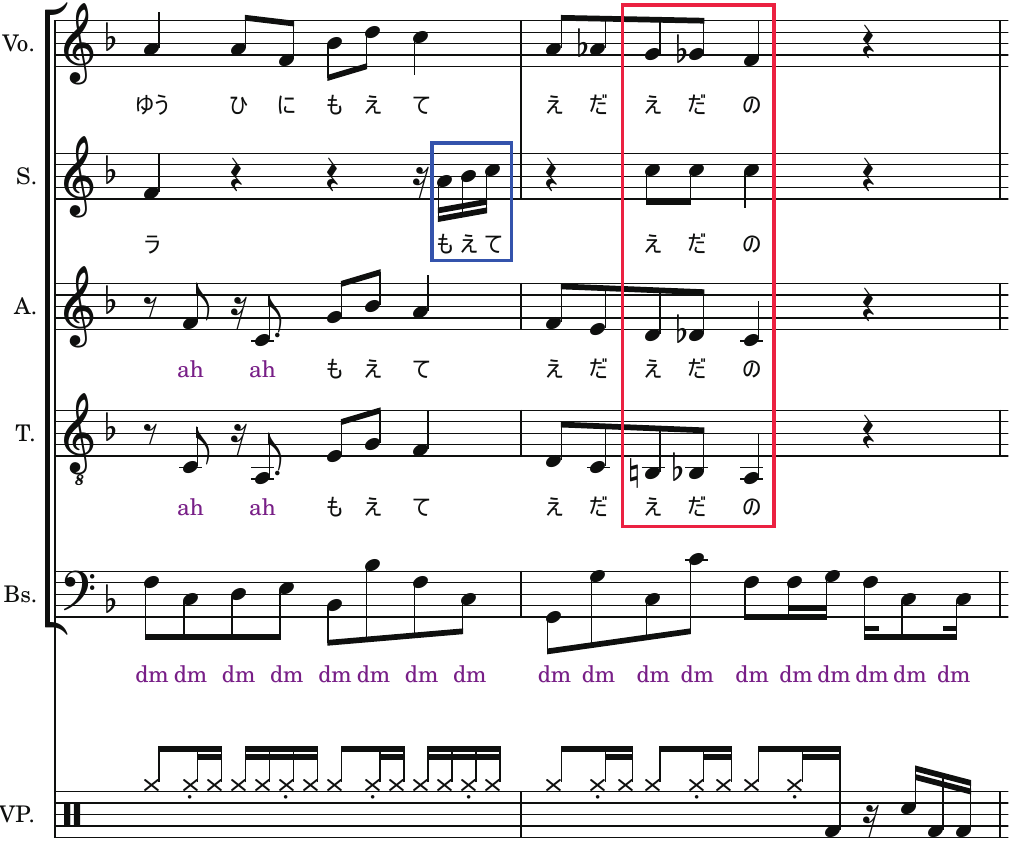}
    \caption{Excerpt of musical score of ``Poplar'' in our jaCappella corpus.}
    \label{fig:score_example}
\end{figure}
In accordance with our design strategy (see Section~\ref{sec:design}), we arranged 35 out-of-copyright Japanese children's songs into vocal ensemble songs.
The arrangement was performed by five arrangers.
The original children's songs were selected from two books of Japanese children's songs \cite{Nobarasya2010Syoka_en,Nobarasya2010Douyou_en}\footnote{The used songs were categorized into \emph{doyo} and \emph{shoka}. Doyo are well-known Japanese children's songs and shoka are Japanese children's songs taught in school.}, which include musical scores of melodies with lyrics.
The Vo parts of our songs were arranged in accordance with the musical scores and the other parts were newly composed.
The lyrics were given for all parts except for the VP part.

The 35 songs are divided into seven subsets (five songs per subset).
The six subsets are named \emph{jazz, punk rock, bossa nova, popular, reggae,} and \emph{enka}.
The songs in each subset were arranged to the corresponding music genre.
For example, the Bs parts of the jazz subset have sequences of equal-duration notes whose pitches alternately go up and down (a.k.a. walking bass lines).
This is one of the typical features of jazz music.
The remaining subset, named \emph{neutral}, consists of the songs that were arranged to retain the mood of the original songs.

We created musical scores of the songs in the MusicXML format~\cite{musicxmlbook}.
This format can include various musical symbols and lyrics and is widely used in MIR studies~\cite{mullerbook} and singing voice synthesis~\cite{Hono2021IEEEACMTASLP}.
Fig.~\ref{fig:score_example} shows an example of the created musical scores.
The created songs include sections where multiple singers synchronously sing the same lyrics (the region surrounded by red lines) and sections where part of singers sing asynchronously with the other singers (the region surrounded by the blue lines).
Our songs well capture these characteristics that frequently appear in the commercial music collection.

\subsection{Recording of singing voices} \label{sec:rec}
\begin{table}[t]
    \centering
    \caption{Singer IDs and average durations of audio recordings in our jaCappella corpus. Singer IDs are comma-separated from left to right for Vo, S, A, T, Bs, and VP, respectively}
    \begin{tabular}{c|cc}
        \toprule
        Subset & Singer IDs & Total dur. [s] \\
        \midrule
        Jazz & Vo1, S1, A1, T1, Bs1, VP1 & 226.7 \\
        Punk rock & Vo2, S2, A2, T2, Bs1, VP2 & 310.7 \\
        Bossa nova & Vo3, S3, A2, T3, Bs2, VP3 & 334.5 \\
        Popular & Vo1, S1, A1, T1, Bs1, VP1 & 352.5 \\
        Reggae & Vo3, S3, A2, T3, Bs2, VP3 & 228.7 \\
        Enka &Vo2, S2, A2, T2, Bs1, VP2 & 361.1 \\
        Neutral &Vo1, S4, A3, T4, Bs1, VP4 & 260.1 \\
        \bottomrule
    \end{tabular}
    \label{tab:audio_statistics}
\end{table}
We conducted recordings of singing voices along with the vocal ensemble songs.
The recordings were performed per singer to avoid COVID-19 infection.
The singing voices were recorded with Shure SM58 microphones in recording studios at a sampling frequency of 48 kHz.
We employed 20 Japanese semi-professional singers: three singers for Vo, four singers for S, three singers for A, four singers for T, two singers for Bs, and four singers for VP.
The Vo, S, and A singers were female and the T, Bs, and VP singers were male.
The vocal ensembles were assembled per subset.
Table~\ref{tab:audio_statistics} shows the singer identifiers (IDs) and total durations of the songs per subset.

After the recordings, we created the best take by combining several takes as in an usual music production process.
The pitch corrections were applied to these signals when the pitches of the singing voices differed from those of the musical scores. We did not use effects such as reverb or an equalizer often used in a commercial music production process.
All voice parts in each song were time-aligned, which makes it easier to use the data for vocal ensemble separation and synthesis.
We also provided the mixtures of all voice parts.
The resultant audio signals were saved as monaural audio files in the RIFF WAVE format with 24-bit linear quantization.

\begin{figure}[t]
    \centering
    \subfloat[Vo]{
        \includegraphics[width=0.45\columnwidth]{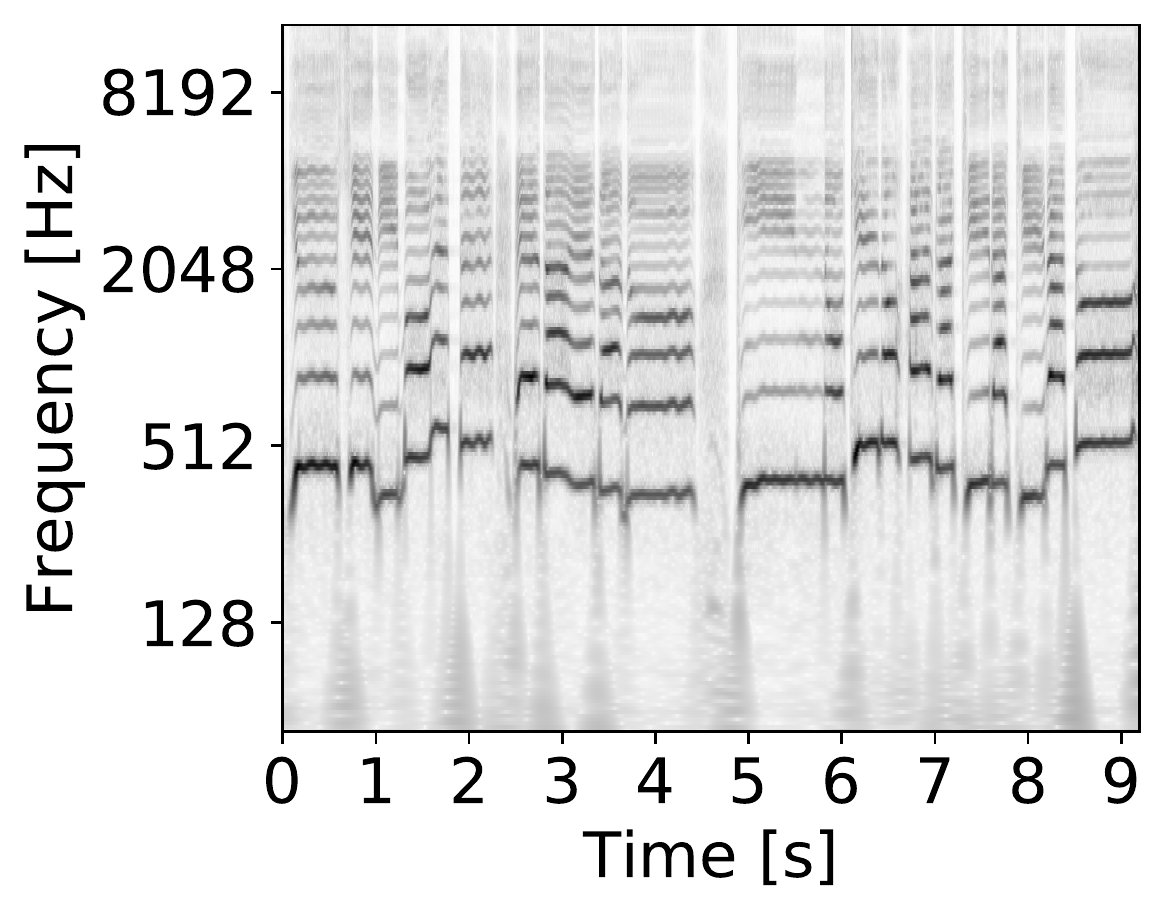}
    }
    \subfloat[S]{
        \includegraphics[width=0.45\columnwidth]{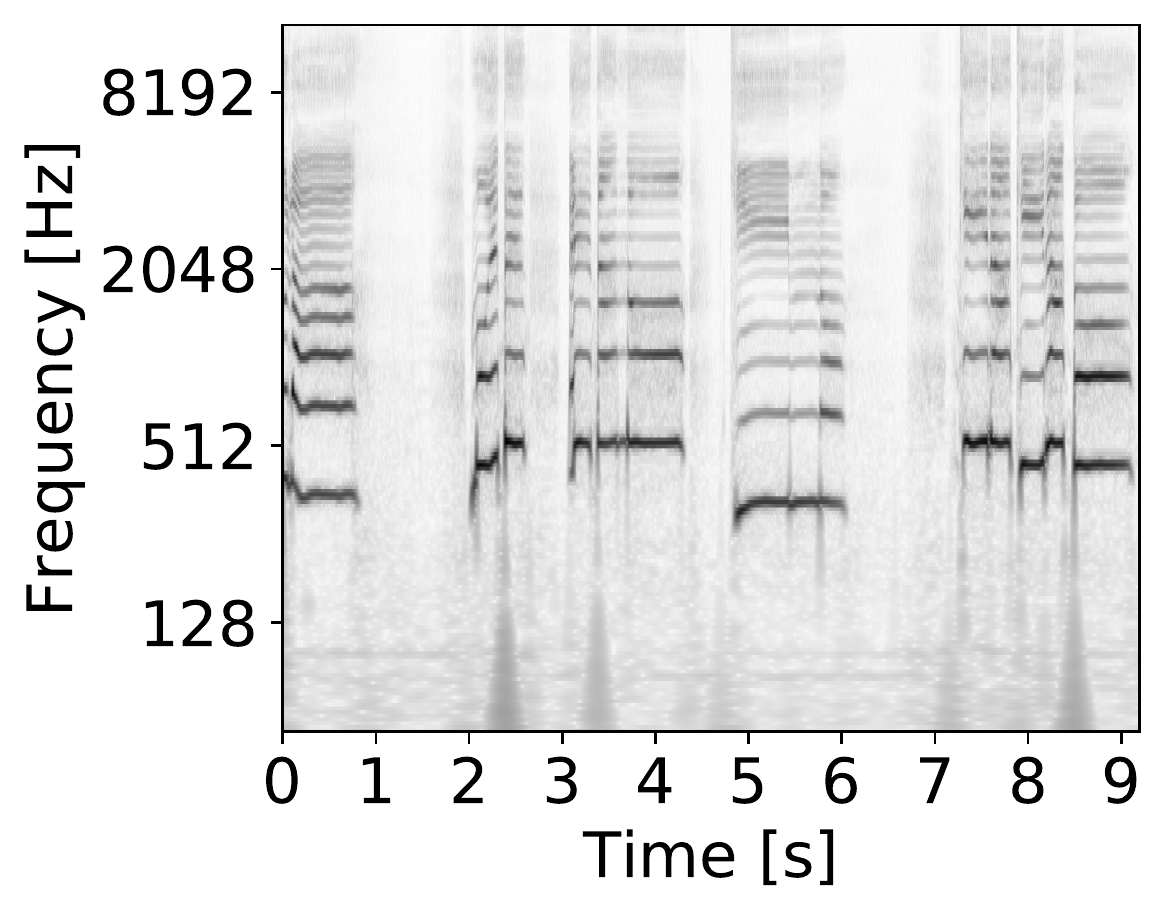}
    }
    \vspace{-2mm}
    \\
    \subfloat[A]{
        \includegraphics[width=0.45\columnwidth]{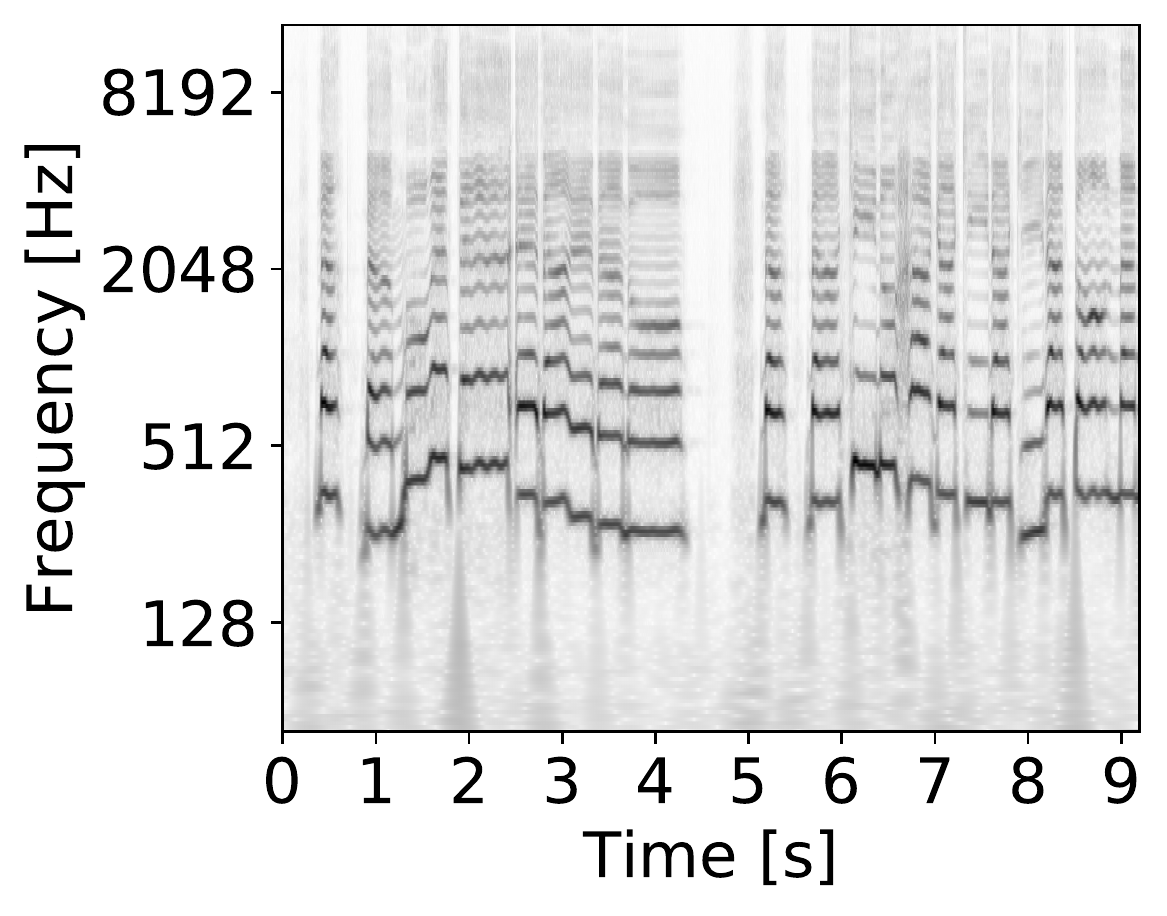}
    }
    \subfloat[T]{
        \includegraphics[width=0.45\columnwidth]{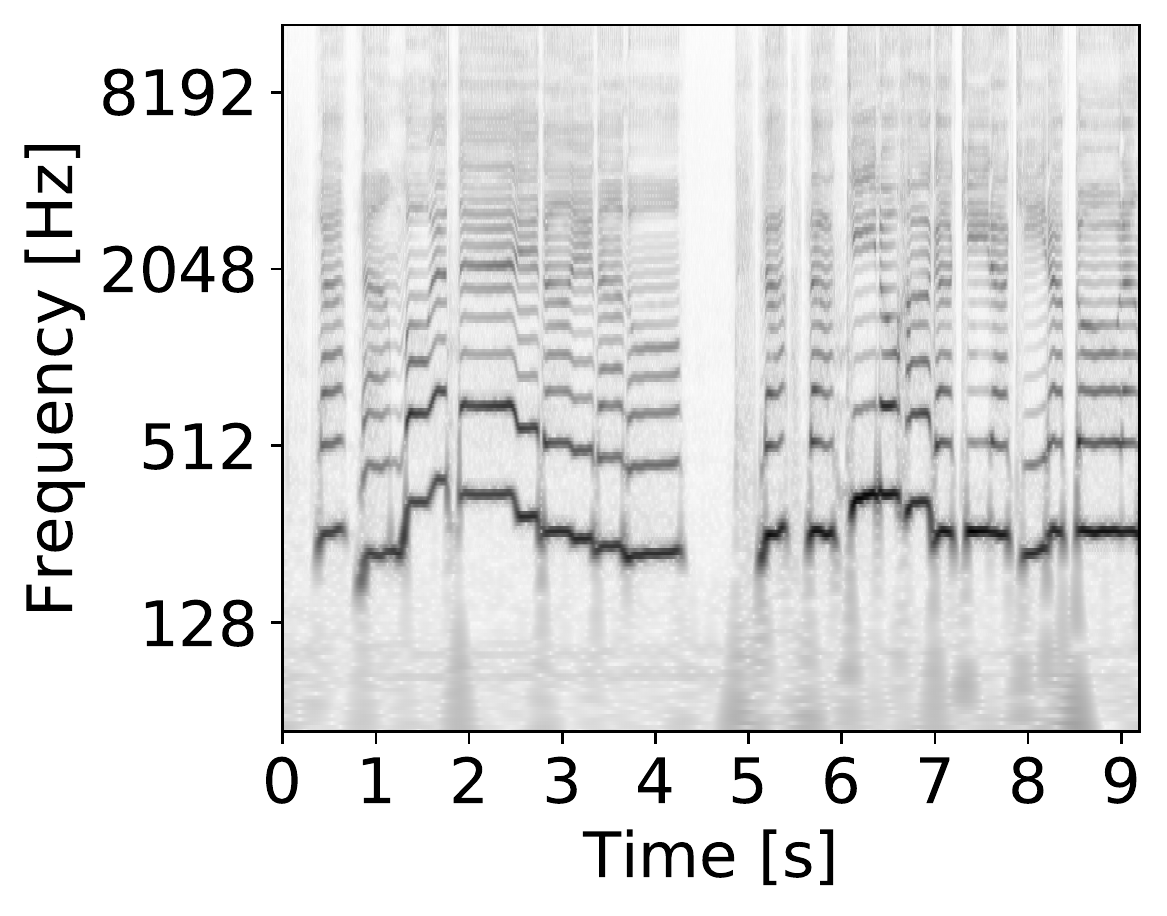}
    }
    \vspace{-2mm}
    \\
    \subfloat[Bs]{
        \includegraphics[width=0.45\columnwidth]{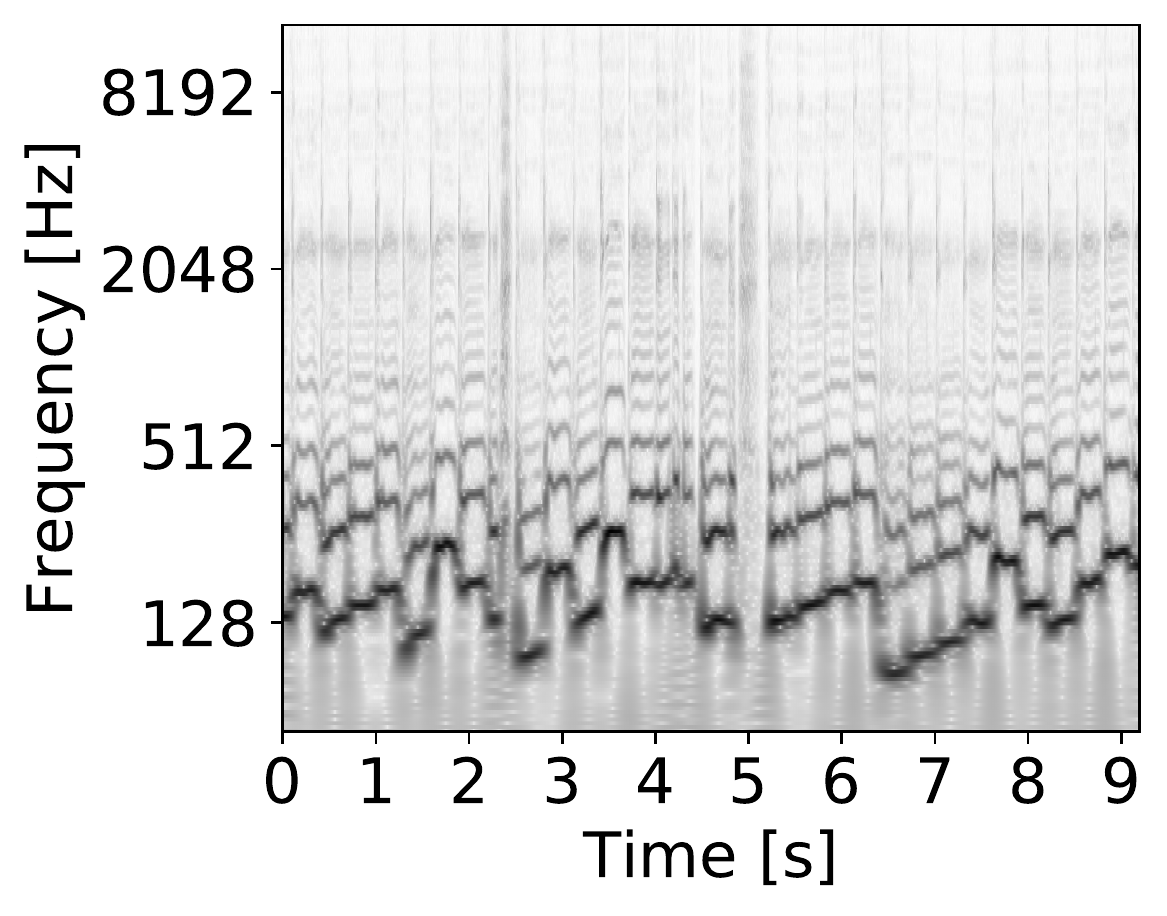}
    }
    \subfloat[VP]{
        \includegraphics[width=0.45\columnwidth]{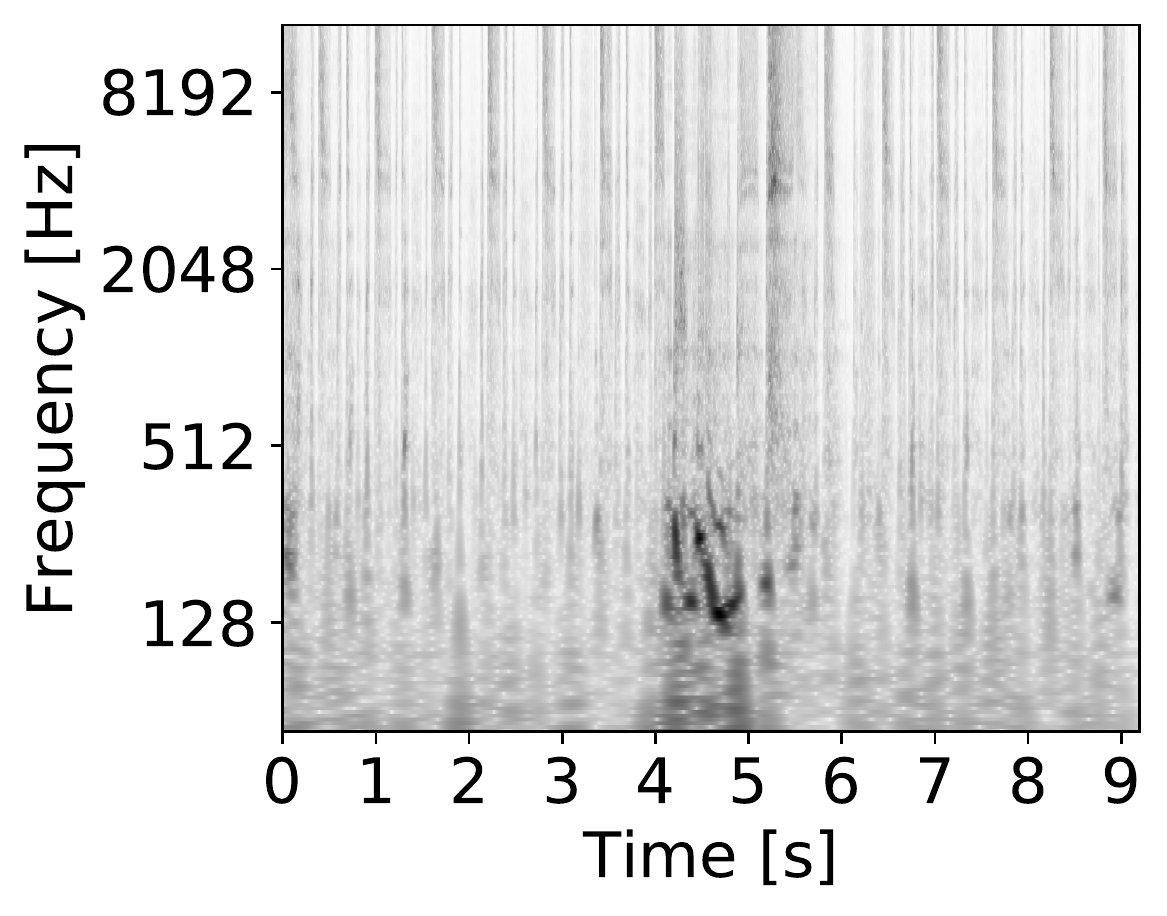}
    }
    \vspace{-2mm}
    \caption{Spectrogram examples of all voice parts.}
    \label{fig:spectrograms}
\end{figure}

Fig.~\ref{fig:spectrograms} show examples of spectrograms of all voice parts.
We can find that the energies of S, A, T, and Bs are distributed in partly overlapped pitch ranges and they change synchronously in part.
The spectrogram of VP has percussive characteristics, but the time-frequency components similar to a voiced spectrogram appear in the range of around 4 to 5 s.
This observation shows that VP has intermediate characteristics between percussive sounds and voices.

\subsection{Analysis of lexical and non-lexical syllables}
\begin{table}[t]
    \centering
    \caption{Frequency of lexical and non-lexical syllables of our jaCappella corpus and commercial music collection (Commercial)}
    \label{tab:frequency_of_singing_styles}
    \begin{tabular}{c|cc|cc}
        \toprule
        \multirow{2}{*}{Voice part} & \multicolumn{2}{c|}{jaCappella [\%]} & \multicolumn{2}{c}{Commercial [\%]} \\
        & Lexical & Non-lexical & Lexical & Non-lexical\\
        \midrule
        Vo & 85.2 & 14.8 & 91.1 & 8.9 \\
        S & 41.9 & 58.1 & 34.2 & 65.8 \\
        A & 32.9 & 67.1 & 35.1 & 64.9 \\
        T & 33.3 & 66.7 & 36.3 & 63.7 \\
        Bs & 0.4 & 99.6 & 2.9 & 97.1 \\
        \bottomrule
    \end{tabular}
\end{table}
As in the commercial music collection, the created songs include ordinary Japanese syllables and nonsense syllables such as ``ah'' and ``dm'' (purple lyrics in Fig.~\ref{fig:score_example}).
The nonsense syllables appear particularly in S, A, T, and Bs.
To distinguish them, we call the former \emph{the lexical syllables} and the latter \emph{the non-lexical syllables}.

The frequencies of these syllables differ in voice part.
Table~\ref{tab:frequency_of_singing_styles} shows the syllable frequencies of the voice parts in the jaCappella corpus and commercial music collection.
We omitted VP since it does not have lyrics.
The frequency of the lexical syllables is high for Vo and quickly decreases from S to Bs.
Since most of the Bs syllables imitate bass instrument sounds, the frequency of non-lexical syllables (e.g., ``dm'', ``bon'', and ``ban'') is quite high.
These tendencies commonly appear in both collections, showing that our corpus captures biases of real vocal ensemble songs in terms of lexical and non-lexical syllables.
For Vo and S, the jaCappella corpus has more non-lexical syllables than the music commercial collection.
This is because some original songs have short lyrics.
When we created vocal ensemble versions of these original songs, we added non-lexical syllables (e.g., ``fu'' and ``la'') before and after the original lyrics so that the resultant songs were longer than at least 30 s.

\section{Vocal Ensemble Separation Experiment} \label{sec:exp}
\subsection{Experimental settings}
As an example application, we used the jaCappella corpus for a vocal ensemble separation task.
This task aims at separating audio signals of the six voice parts from their monaural mixture.
The sampling frequency was set to 48 kHz.
We chose seven songs for test, one from each subset.
The total duration of the test songs was 350.2 s.
The other songs were used for training.
Note that the separation was done in the singer-closed setting.
We computed average improvements of scale-invariant source-to-distortion ratio (SI-SDR) \cite{LeRoux2019ICASSP} as an evaluation metric.
The input SI-SDRs of the test data ranged from -23 to -1.3 dB, which shows the difficulty of the separation task.

To increase the amount of training data, we used an intersong mixing method proposed in \cite{Defossez202MDX}.
It synthesizes realistic mixtures by mixing audio signals of voice parts of different songs, changing their tempi and pitches adequately.
The augmented data of the 28 training songs were divided into training and validation sets.
The total duration of the training and validation sets were 7650.7 and 2811.7 s, respectively.
We experimentally observed that this data augmentation improved the separation performance.
We also used on-the-fly data augmentations \cite{Sawata2020ArXiv}: random cropping of the training audio segments and random amplification within the range of $[0.25,1.25]$.

\subsection{Compared methods}
We compared three methods: X-UMX \cite{Sawata2020ArXiv}, an adaptation of DPTNet \cite{Chen2020Interspeech} for vocal ensemble separation (DPTNet) \cite{Sarkar2021Interspeech}, and MRDLA \cite{Nakamura2021IEEEACMTASLP}.
X-UMX performs the separation in the spectrogram domain while DPTNet and MRDLA separates mixtures in the waveform domain.
All methods estimated monaural audio signals of voice parts from monaural mixtures.
The first and last layers of the waveform-based methods were modified to have one input channel and six output channels, respectively.
We trained all methods with a NVIDIA A100 GPU and picked up the trained models at the epoch with the lowest validation loss.
The other settings were as follows:

\noindent \textbf{X-UMX:} This method was a baseline method in Music Demixing Challenge 2021~\cite{Mitsufuji2022FSP}.
It was originally proposed for music source separation and we applied it to vocal ensemble separation.
The number of epochs was 1000 and the other parameters were the same as in the official implementation\footnote{\url{https://github.com/asteroid-team/asteroid/tree/master/egs/musdb18/X-UMX}}.

\noindent \textbf{DPTNet:} This method is an adaptation of DPTNet for the separation of four-part vocal ensembles.
We set the kernel sizes and strides of the first convolutional and last transposed convolutional layers to 32 and 16, respectively.
The number of epochs was 600 and the other settings were the same as in the official implementation\footnote{\url{https://github.com/saurjya/asteroid/tree/4e00daa4c4da77bbee6c0109fa4e2c3611217e72}}.
Since this method uses permutation invariant training \cite{Yu2017ICASSP}, the SI-SDRs of this method were computed after the best permutations were obtained.

\noindent \textbf{MRDLA \cite{Nakamura2021IEEEACMTASLP}:} This method is a wavelet-based extension of Wave-U-Net \cite{Stoller2018ISMIR}.
It was originally proposed for music source separation and we adapted it to vocal ensemble separation.
We adopted the network architecture with the Haar wavelet, which achieved nearly best separation performance in \cite{Nakamura2021IEEEACMTASLP} and is easy to implement.
We applied it the following modifications.
The number of channels of all layers were increased. More specifically, with the notation used in \cite{Nakamura2021IEEEACMTASLP}, $C^{\text{(e)}}$ increased from 18 to 32.
The kernel sizes of convolutional layers were increased to 21 and parametric rectified linear units were replaced with Gaussian error linear units.
We used a loss function that combines time-domain and time-frequency-domain loss functions \cite{Kong2022ICASSP}.
These modifications greatly improved the separation performance.
The number of epochs was 1000 and the other settings were the same as X-UMX.

The codes used are publicly available at \url{https://github.com/TomohikoNakamura/asteroid_jaCappella}.

\subsection{Results}
Table~\ref{tab:results} shows average SI-SDR improvements per voice part.
Compared with X-UMX, the waveform-based methods (DPTNet and MRDLA) provided higher separation performance.
This result indicates that the waveform-based methods are suitable for vocal ensemble separation.
DPTNet provided highest average SI-SDRs for Vo, T, and Bs.
MRDLA achieved highest average SI-SDRs for the other voice parts.
Their average performances were similar for most of the voice parts, but DPTNet clearly outperforms MRDLA for Bs.
The VP sounds were well separated in all methods, showing that the percussive characteristics of VP spectrograms make the separation easier.
This result is reminiscent of the fact that drum sounds tend to be well separated in music source separation of vocals, bass, drums, and other instruments \cite{Mitsufuji2022FSP}.

Some separated audio signals are available at \url{https://tomohikonakamura.github.io/Tomohiko-Nakamura/demo/jaCappella_sep}.
When listening to them, we can find that the separated signals of X-UMX (particularly for Bs) tend to lack high frequency components and those of the waveform-based methods tend to include high frequency noises.
A similar tendency was observed in music source separation \cite{Schaffer2022arXiv}.
The separated signals of DPTNet included high frequency noises in all voice parts.
MRDLA provided clearer but sometimes choppy separation results.
For all methods, the separated signals of Vo, S, A, and T were often contaminated with the other voice parts, sometimes with higher energies than the target sources.
This trend is common in all methods, showing the difficulty of this separation task.
Compared with these voice parts, the separated signals of Bs and VP had less leakage from the other voice parts.
This should be because Bs was in the much lower pitch range and the timbre of VP greatly differs from the other voice parts as mentioned in Section~\ref{sec:rec}.
These results clearly show that the jaCappella corpus is a challenging resource to develop and test vocal ensemble separation methods.

\begin{table}[t]
    \centering
    \caption{Average SI-SDR improvements of separation methods}
    \label{tab:results}
    \begin{tabular}{c|cccccc}
        \toprule
        \multirow{2}{*}{Method} & \multicolumn{6}{c}{Voice part [dB]} \\
        & Vo & S & A & T & Bs & VP \\
        \midrule
        X-UMX &         7.5 &         10.7 &          13.5 &   10.2 &   9.1 &  21.0 \\
        DPTNet&\textbf{8.9} &          8.5 &          11.9 &   \textbf{14.9} &  \textbf{19.7} & 21.9 \\
        MRDLA &         8.7 & \textbf{11.8}& \textbf{14.7} &            11.3 &          10.2 &  \textbf{22.1} \\
        \bottomrule
    \end{tabular}
\end{table}

\section{Conclusion}
We constructed the jaCappella corpus that consists of 35 Japanese vocal ensemble songs and separate audio recordings of individual voice parts.
The created songs are vocal ensemble arrangements of Japanese children's songs.
Unlike conventional vocal ensemble datasets, their voice parts include Vo and VP and the arrangement cover various music genres.
We confirmed that our corpus has distributions of lexical and non-lexical syllable similar to commercial vocal ensemble songs.
Experimental evaluation demonstrated that the jaCappella corpus is a challenging resource to develop and test algorithms of vocal ensemble separation.
We believe that our corpus accelerates MIR research on vocal ensembles.

\vfill
\pagebreak

\bibliographystyle{IEEEbib}
\bibliography{refs}

\end{document}